
\def\dag{\dagger}
\def\del{\partial}

\def\a{\alpha}     
\def\b{\beta}      
     
\def\d{\delta}

\def\n{\nu}

\def\br{\langle}
\def\ke{\rangle}
\def\ve{\vert}

\def\Winf{$W_{\infty}\  $}
\def\kvecz{\ve z_1  \cdots z_N \ke}

\def\zbar{\bar{z}}

\def\half{{\textstyle{1 \over 2}}}

\headline={\ifnum\pageno=1\firstheadline\else
\ifodd\pageno\rightheadline \else\leftheadline\fi\fi}
\def\firstheadline{\hfil}
\def\rightheadline{\hfil}
\def\leftheadline{\hfil}
        \footline={\ifnum\pageno=1\firstfootline\else\otherfootline\fi}
\def\firstfootline{\rm\hss\folio\hss}
\def\otherfootline{\hfil}

\font\tenrm=cmr10

\font\elevenbf=cmbx10 scaled\magstep 1
\font\elevenrm=cmr10 scaled\magstep 1
\font\elevenit=cmti10 scaled\magstep 1

\font\ninerm=cmr9

\nopagenumbers
\hsize=6.0truein
\vsize=8.5truein
\parindent=1.5pc
\baselineskip=10pt
\line{\hfil \ninerm IASSNS-HEP-94/93 }
\medskip
\centerline{\elevenbf ${\bf W_{\infty}}$ ALGEBRAS AND INCOMPRESSIBILITY}
\vglue 7pt
\centerline{\elevenbf IN THE QUANTUM HALL EFFECT {\ninerm \footnote{*}
{\baselineskip-11pt
Talk presented in Mt. Sorak Summer School, S. Korea, June 27-July 2,
1994
and XXth Int. Colloquium on Group Theoretical Methods in Physics,
Osaka, Japan, July 4- July 9, 1994, to appear in proceedings.}}}
\vglue 1.0cm
\centerline{\elevenrm DIMITRA KARABALI {\ninerm \footnote{$\dag$} {
\baselineskip=11pt
This work was supported in part by the DOE grants DE-FG02-85ER40231 and
DE-FG02-90ER40542}}
}
\baselineskip=13pt
\centerline{\elevenit Institute for Advanced Study}
\baselineskip=12pt
\centerline{\elevenit Princeton, NJ 08540, USA}

\vglue 0.8cm
\centerline{\tenrm ABSTRACT}
\vglue 0.3cm
  {\rightskip=3pc
 \leftskip=3pc
 \tenrm\baselineskip=12pt
 \noindent
We discuss how a large class of incompressible quantum Hall states can be
characterized as highest weight states of
different representations of the \Winf algebra. Second quantized expressions
of the \Winf generators are explicitly derived in the cases of
multilayer Hall states, the states proposed by Jain to explain the
hierarchical filling fractions and the ones related by particle-hole
conjugation.
\vglue 0.8cm }
\baselineskip=14pt
\elevenrm
The study of planar, charged nonrelativistic fermions in a strong magnetic
field finds important applications in condensed matter problems, such as the
quantum Hall effect (QHE)$^{1,2}$. Such systems have a further, less obvious
connection
to (1+1)-dimensional problems, such as the $c=1$ string model$^{3,4}$. In this
talk I will outline the emergence of an infinite
dimensional algebraic structure, the \Winf algebra, and its role in the integer
and fractional QHE (IQHE and FQHE), based mainly on work presented in refs. 6
and 7.
\vglue 0.6cm
\line{\elevenbf 1. \Winf algebras for IQHE \hfill}
\vglue 0.4cm
As is well known, the spectrum of planar, nonrelativistic fermions in the
presence of a transverse, uniform magnetic
field $B$, consists of infinitely degenerate levels, the so-called Landau
levels. The energy gap between adjacent Landau levels is
$\omega = B/M$, where $M$ is the fermionic mass ($\hbar=c=e=1$). For large
$B$ we can consider
the fermions restricted to the lowest Landau level (LLL). We further assume
that
$B$ is sufficiently strong to align all electronic spins, so we
can neglect the spin degree of freedom.
In the symmetric gauge $\vec{A} (\vec{x}) = {\textstyle{B \over 2}} (y,-x)$,
the LLL condition can be written as
$$ (\del _z + \half \zbar) ~ \Psi (\vec{x}) = 0 \eqno (1.1) $$
where $z= \sqrt{\textstyle {B \over 2}} (x+iy),~
\zbar= \sqrt{\textstyle {B \over 2}} (x-iy)$. The LLL wavefunctions
are of the form
$$ \Psi (\vec{x}) = f(\zbar) \exp \left( -{\textstyle {1 \over 2}} |z| ^2
\right) \eqno (1.2)$$
where $f(\zbar)$ is a polynomial in $\zbar$. Upto an exponential
factor which can be absorbed in the definition of the measure, the LLL
wavefunctions depend
only on the antiholomorphic variables $\zbar$, while the holomorphic ones
become essentially the canonical momenta ($z \rightarrow \del_{\zbar}$ after
taking into account the appropriate ordering)$^{8}$. This reflects the fact
that
the original coordinate space of electrons constrained in the LLL becomes
the phase space of a one-dimensional system$^{3}$.

In a second quantized language the LLL condition can be promoted to an operator
equation and the corresponding LLL fermion operator has the form
$$
\Psi (\vec{x} ,t) ={\sqrt{B \over {2 \pi}}} e^{- \half |z|^2}
\sum _{l =0}^{\infty}  C_l (t) {{\bar{z}^{l}} \over {\sqrt{l!}}}
\equiv {\sqrt{B \over {2 \pi}}}  e^{- \half |z|^2}  \psi (\zbar,t)
\eqno(1.3) $$
where $C_l$'s are operators which annihilate fermions of angular momentum $l$
and satisfy the usual anticommutation relations
$ \{ C_{l}^{\dag} , C_{l'} \} =  \d _{l,l'}$.

In the absence of an external potential and interactions there is an infinite
degeneracy with respect to angular momentum, so the system is symmetric under
independent unitary transformations in the space of $C$'s:
$$
{C}_l (t)= u_{lk} {C}_k (t)=\br{l} \ve u \ve k \ke {C}_k (t) \eqno (1.4)
$$
The corresponding
infinitesimal transformation for the LLL fermion operator is
$$
\d {\Psi}^I (\vec{x},t) = i\ddag \xi ( \del_{\zbar}+{z\over 2} ,
\zbar   )\ddag ~
{\Psi}  (\vec{x} ,t)   \eqno (1.5)
$$
where $\xi (z,\zbar)$ is a real function
and $\ddag\ \ \ \ddag$ indicates that the operators $ \del_{\zbar}+{z\over
2}$ act from the left.
These transformations preserve the LLL condition and the particle
number, i.e.,
$ \int d {\vec{x}} \d \rho (\vec{x},t) =0$,
where
$\rho  (\vec{x},t)
=  \Psi ^{ \dag}(\vec{x},t) \Psi  (\vec{x},t)$ is the LLL fermion density.
The corresponding generators are given by$^{4-7}$
$$
\rho [{\xi}] \equiv
\int  d^2 z e^{-|z|^2} \psi ^{\dag } (z)
{}~ \ddag\xi ( \partial_{\bar{z}} , \bar{z} )\ddag ~
\psi  (\bar{z}) \eqno (1.6) $$
where $d^2 z \equiv {B \over {2 \pi}} dx dy$, and they
satisfy an infinite dimensional algebra given by
$$ \eqalignno{
& [\,\rho  [\,\xi_1 \, ],\rho  [\, \xi_2 \, ]\, ]= {i\over B}\rho
[\{\!\!\{\xi_1 ,\xi_2 \}\!\!\}]  &{}\cr
& \{\!\!\{\xi_1 ,\xi_2 \}\!\!\}=iB{\sum_{n=1} ^{\infty}}{{(-)^n}\over{n!}}
\left(
{\partial_{z} ^{n}}\xi_1 {\partial_{\bar{z}} ^n}\xi_2 -
{\partial_{\bar{z}} ^{n}}\xi_1 {\partial_{z} ^n}\xi_2\right) &(1.7) \cr}
$$
$\{\!\!\{ \}\!\!\}$ is the so-called Moyal bracket. This is the $W _{\infty}$
algebra$^{9}$.

In the absence of an external potential and interactions, the \Winf algebra
corresponds to a symmetry of the problem. In realistic
situations the electrons are confined in a finite region. In the infinite plane
geometry this can be achieved by introducing an external confining potential
$V(
\vec{x})$,
for example a central harmonic oscillator potential. Such a term spoils the
infinite degeneracy with respect to angular momentum by assigning higher
energy to higher angular momentum states, but the resulting Hamiltonian,
$H = \int d^2 x V(\vec{x}) \rho (\vec{x}, t) $,
is a member of the \Winf algebra$^{4,5}$. In this case the
\Winf algebra does not correspond to a symmetry anymore; instead it provides a
spectrum generating algebra. In order to illuminate this role of the \Winf
algebra we consider the action of the \Winf generators on the many-body
ground state of electrons filling up the first Landau level, i.e. $\nu =1$
(where $\nu$ is the filling fraction, the ratio between the number of
electrons and the degeneracy of the Landau level). The $\nu =1$ ground state is
$$|\Psi_{\n=1} \ke _0 = C^{\dag} _{0} ... C^{\dag} _{N-1}  |0 \ke \eqno (1.8)
$$
for $ N $ electrons. This forms an incompressible, circular droplet of radius
$\sim \sqrt{N/B}$ and uniform density $\rho = B / 2 \pi$.
Compression corresponds to lowering the
angular momentum, but since all available states in the LLL are occupied
($\nu =1$), that would require an electron to jump to a higher Landau
level. However, for a large magnetic field, this is not energetically allowed
due to the big energy gap.
On the other hand, deformations that would result in transitions to states
with higher angular
momentum are allowed and cost some energy due to the confining potential. These
excitations can be generated by the action
of \Winf generators on the ground state.
Inspection of the \Winf generators in the basis
$\xi (z,\bar{z})=z^l \bar{z}^k$ shows that the operators
$\rho _{lk}$ decrease the angular
momentum for $l>k$ and increase the angular momentum for
$l<k$, where
$\rho _{lk} \equiv \int d^2 z e^{-|z|^2} \psi ^{\dag} (z)
(\del_{\zbar}) ^l (\zbar) ^k \psi  (\zbar)$. Thus we find
that$^{5-7}$
$$ \eqalignno{
& \rho _{lk} |\Psi_{\n=1} \ke _0=0 ~~~~~~~~~~~~  {\rm if}~~l>k &{}\cr
& \rho _{lk} |\Psi_{\n=1} \ke _0=|\Psi > ~~~~~~    {\rm if}~~l\le k
&(1.9) \cr}
$$
where $|\Psi  \ke $ corresponds to excitations of higher angular momentum.

The first line in Eq. (1.9) provides an  algebraic statement for the
incompressibility of the ground state, by characterizing the
ground state as the highest weight state
of the \Winf algebra$^{5}$. In fact we shall show that even for more general
filling fractions the incompressibility of the corresponding ground states
can be algebraically expressed as the ground state being the highest weight
state of a \Winf algebra.

Before generalizing to other filling fractions I would like to briefly mention
the relation between the \Winf algebra and the algebra of area preserving
diffeomorphisms. This can be easily understood in the case of LLL fermions.
\Winf transformations were earlier introduced as unitary transformations.
Their
classical analogue, therefore, are canonical transformations
which preserve the area element of the phase space. Given that the LLL phase
space corresponds to the original two-dimensional
coordinate space of the system$^{3}$, as mentioned earlier, these canonical
transformations are the
area preserving diffeomorphisms. In terms of excitations one can understand
the relation between the two algebraic structures by considering the edge
excitations$^{10}$, i.e., $k-l \sim O(1)$. These low energy excitations
correspond to boundary fluctuations of the ground state droplet and
can be described by one-dimensional chiral boson (fermion) fields. In terms
of these, and
upon restriction to the edge excitations, one can show
that the original \Winf algebra
reduces to the algebra of area-preserving
diffeomorphisms$^{4,11}$.

The previous analysis can be easily extended to the case where the fermions
fill up the first $n$ Landau levels, $\nu =n$.  A simple
analysis of the Hamiltonian and the corresponding single-body energy and
angular momentum wavefunctions shows that the fermion operator can
be now expanded as
$$
\Psi (\vec{x} ,t) ={\sqrt{B \over {2 \pi}}} e^{- \half |z|^2}  \sum _{I=0} ^{n}
\sum _{l =0}
^{\infty} i^I C_l ^I (t)
{{(z-\partial_{\bar{z}})^I} \over {\sqrt{I!}}} {{\bar{z}^{l}} \over
{\sqrt{l!}}}
\equiv {\sqrt{B \over {2 \pi}}}  e^{- \half |z|^2} \sum _{I=0} ^{n} \psi ^I
(z, \zbar,t)
\eqno(1.10) $$
where $I$ indicates the Landau level and is related to the energy and $l-I$
measures
the angular momentum.
The operators $C_l ^I$ satisfy
$ \{ C_{l}^{\dag I} , C_{l'}^{I'} \} = \d _{I,I'} \d _{l,l'}$.
There are now $n$ mutually commuting \Winf generators corresponding to
independent unitary
transformations acting at each Landau level. They are of the form
$$
\rho^{I} [{\xi}] \equiv
\int  d^2 z e^{-|z|^2} \psi ^{I \dag } (z,\zbar)
{}~ \ddag\xi ( \partial_{\bar{z}} , \bar{z}-\del _z )\ddag ~
\psi^I  (z,\bar{z}) ~~~~~~~~I=0,1,..,n  \eqno (1.11) $$
Their action on the $\nu =n$ ground state $|\Psi_{\n=n} \ke _0$, where
(for $N'=nN$ electrons)
$$|\Psi_{\n=n} \ke _0 = \prod _{I=0} ^{n-1} (C^{\dag I} _{0} ... C^{\dag I}_
{N-1} ) |0 \ke ,
\eqno (1.12) $$
is quite similar to Eq. (1.9). We find that$^{5,7}$
$$ \eqalignno{
& \rho^I _{lk} |\Psi^I_{\n=n} \ke _0=0 ~~~~~~~~~~~~  {\rm if}~~l>k
{}~~~~~~I=0,1,...,n-1 &{}\cr
& \rho^I _{lk} |\Psi^I_{\n=n} \ke _0=|\Psi_I > ~~~~~~    {\rm if}~~l\le k
{}~~~~~I=0,1,...,n-1 &(1.13) \cr}
$$
where $\rho^I _{lk} \equiv \int d^2 z e^{-|z|^2} \psi ^{I \dag} (z,\zbar)
(\del_{\zbar}) ^l (\zbar-\del _z) ^k \psi  (z,\zbar)$ and $ |\Psi _I \ke $
corresponds
to excitations of higher angular momentum at the $I$-th level.
As in the $\nu=1$ case, the incompressibility of the $\nu =n$ ground state is
algebraically expressed as the ground state being the highest weight state of
a \Winf algebra. In the next section we shall seek the generalization of
this to the fractional quantum Hall states.

\vglue 0.6cm
\line{\elevenbf 2. \Winf algebras for FQHE \hfill}
\vglue 0.4cm
The main experimental feature of both the IQHE and FQHE, namely the appearance
of a series of plateaux where the Hall conductivity is quantized and
proportional
to the filling fraction $\nu$, while the longitudinal conductivity vanishes, is
attributed to the existence of a gap, which gives rise to an incompressible
ground
state. For IQHE, the essential physics can be well understood in
terms of noninteracting fermions. The energy gap is the cyclotron energy
separating
adjacent Landau levels and for a large magnetic field the Coulomb interaction
can
be neglected. The noninteracting picture is nonapplicable in the case of the
FQHE,
where the Coulomb interaction
among electrons is important in producing an energy gap. Much of our
understanding of the FQHE relies on successful trial wavefunctions, such as
the Laughlin wavefunctions$^{12,13}$ and the ones proposed more recently by
Jain$^{14}$.
In both cases they
correspond to incompressible configurations of uniform density
$\rho = \nu B/2 \pi$.

Earlier, in the case of the IQHE, we have seen that the incompressibility of
the ground state is closely related to the existence of the \Winf algebra
structure.
This relation can be extended to the FQHE$^{6,7,14,15}$.
I shall first describe the derivation of \Winf algebras for
$\n =1/m$ Laughlin states and their relation to $\n =1$ \Winf
algebras which
will be crucial in constructing similar algebraic structures for quantum Hall
fluids of general filling fraction. Here we consider the electrons confined
in the lowest Landau level.

The main point in this derivation is the simple observation that the $\nu =1/m$
Laughlin ground state wavefunction is related to the $\n =1$ wavefunction by
attaching $2p$ (where $m=2p+1$) flux quanta to each electron
$$
\Psi ^0 _{\n =1/m} = \prod _{i<j} (\zbar _{i} - \zbar _{j}) ^{2p} \Psi ^0 _{\n
=1} \eqno (2.1)
$$
Based on this we find that the \Winf generators for $\nu =1/m$ are related to
the $\nu =1$ generators by a similarity transformation. The corresponding
second
quantized expression can be written in terms of fermion and quasihole
operators as$^{6}$
$$W_{2p}[\xi]= \int d^2 z e^{-|z|^2} \psi^{\dag} (z) e^{2p \alpha
(\bar{z})} ~ \ddag \xi ( \partial _{\bar{z}}, \bar{z}) \ddag ~ e^{-2p \alpha
(\bar{z})} \psi(\bar{z}) \eqno(2.2)
$$
where $\alpha (\bar{z})=\int d^2 z' e^{-|z'|^2} \ln (\bar{z}-\bar{z}')
\psi^{\dag}(z') \psi (\bar{z}') $
and $e^{\a (\zbar)}$ is the quasihole operator since
$$
e^{\alpha (\zbar )}| \Psi \ke = \int d^2 z_1...d^2 z_N
e^{-\sum_{i}|z_i |^2}  F(\zbar _1,...,\zbar _N)
\prod_i (\zbar -\zbar _i )\kvecz \eqno (2.3)
$$
with
$| \Psi \ke = \int d^2 z_1...d^2 z_N
e^{-\sum_{i}|z_i |^2} F(\zbar _1,...,\zbar _N) \kvecz  $.

The operators $W_{2p}$
satisfy a strong \Winf algebra$^{6,7}$
$$[W_{2p}[\xi _1], W_{2p}[\xi _2]] = W_{2p}[\{\!\!\{\xi _1 ,
\xi _2 \}\!\!\}] \eqno(2.4)
$$
They further play the role of a spectrum generating algebra
in the space of Laughlin states and in particular the ground state is the
highest weight state
$$ (W_{2p})_{lk} |\Psi_{\n=1/m} \ke _0=0 ~~~~~~~~~~~~  {\rm if}~~l>k
\eqno (2.5)
$$
This provides an algebraic statement of incompressibility for the Laughlin
ground states.
The operators $W_{2p}$ form a one-parameter
family of \Winf representations.

These ideas can be now extended$^{7}$ to include other incompressible
states corresponding to filling fractions $\n \ne 1/m$.
In general all these states can be characterized as the highest weight states
of different realizations of a \Winf algebra.
I shall briefly present
three such cases and explicitly write down the second quantized expressions
of the corresponding \Winf generators
1) $\nu =1-1/m$ states 2) multilayer systems and 3) Jain states.
\vglue 0.4cm
\line{\elevenit 2.1. $\nu = 1- 1/m$ states \hfill}
\vglue 0.1 cm
Using the idea of particle-hole conjugation$^{17}$, we can write the
$\n =1 -{1 \over m}$ ground state, in the thermodynamic limit and up to
normalization factors, as
$$
|\Psi _{\n =1 - 1/m} \ke _0 \sim  \int d^2 z_1 ... d^2 z_M e^{-\sum |z_i|^2}
\prod
_{i<j} (z_i
- z_j) ^m \psi (\zbar _1) ... \psi (\zbar _M) |\Psi _{\n =1} \ke _0 \eqno (2.6)
$$
Introducing the operator $\b (z)
=\int d^2 z' e^{-|z'|^2} \ln (z-z')
\psi (\zbar ') \psi ^{\dag} (z') $,
we find that
$$
\tilde{W} _{2p}[\xi]= \int d^2 z e^{-|z|^2} \psi (\zbar) e^{2p \b
(z)} ~ \ddag \xi ( \partial _{z}, z) \ddag ~ e^{-2p \b
(z)} \psi ^{\dag} (z) \eqno(2.7)
$$
satisfy a \Winf algebra and the state in Eq. (2.6) satisfies a highest weight
condition.
The operator $\tilde{W} _{2p}$ is essentially the charge-conjugated version of
$W _{2p}$.
\vglue 0.4cm
\line{\elevenit 2.2. Multilayer states \hfill}
\vglue 0.1 cm
In order to obtain more general filling fractions, one may consider systems
where several distinct species of electrons are involved. Wavefunctions of the
form
$$
\Psi ^K (\vec{x} _i ^I) = \prod _{I=1}^{r} \prod _{i<j} (\zbar _i ^I -
\zbar _j ^I)
^{K_{II}} \prod_{I<J} \prod _{i,j} (\zbar _i ^I - \zbar _j ^J) ^{K_{IJ}}
e^{-1/2 \sum _{iI} |z_i|^2}
\eqno (2.8)
$$
have been suggested$^{18}$ as candidates for describing incompressible Hall
states for
an $r$-layer system. $K_{II}$ are odd integers so that the wavefunctions are
antisymmetric under exchange of identical fermions.
The corresponding
filling fraction is given by  $\n = \sum _{I,J} (K^{-1})_{IJ}$ where $K$ is an
$r \times r$ symmetric matrix.
In order to identify the \Winf algebra structure associated with
Eq. (2.8) we introduce $r$ independent lowest Landau level fermion operators
$\psi ^I
(\zbar,t)$
and the corresponding quasihole operators $e^{\a ^{I} (\zbar)}$.
The \Winf generators are now$^{7}$
$$
W_{\tilde{K}} ^I [\xi]= \int d^2 z e^{-|z|^2} \psi^{\dag I} (z)
e^{ \sum _{J} \tilde{K}_{IJ} \alpha ^J
(\bar{z})} ~ \ddag \xi ( \partial _{\bar{z}}, \bar{z}) \ddag ~ e^{- \sum _{j}
\tilde {K} _{IJ} \alpha ^J
(\bar{z})} \psi ^I (\bar{z}) \eqno(2.9)
$$
where $I=1,...,r$ and $\tilde {K} = K -{{\rm I}\!{\rm I}}$
(${{\rm I}\!{\rm I}}$ is the identity matrix). We can show
that the operators $W ^I _{\tilde{K}}$ give rise to $r$ commuting copies of
\Winf algebras and the corresponding ground states
satisfy highest weight conditions.

The \Winf generators of Eq. (2.9) can also be written as
$$
W_{\tilde{K}}^I [\xi]= \int d^2 z e^{-|z|^2} \psi^{\dag I} (z) ~ \ddag \xi
(\partial _{\bar{z}}-\sum _{J} \tilde{K}_{IJ} \int d^2 z' {{\rho ^{J} (z',
\bar{z}')}
\over {(\bar{z}-\bar{z}')}}, \bar{z}) \ddag ~ \psi ^{I}(\bar{z}) \eqno(2.10)
$$
The term $ \sum_{J} \tilde{K} _{IJ} \int d^2 z' {{\rho ^{J} (z',\bar{z}')}
\over
{(\bar{z}-\bar{z}')}}$ plays the role of a gauge potential and it is similar
to the one induced by the Chern-Simons interaction. Analogous first quantized
expressions were also derived in ref.15.
\vglue 0.4cm
\line{\elevenit 2.3. Jain states \hfil}
\vglue 0.1cm
One of the theories proposed to explain the observed fractions for the FQHE is
the one suggested by Jain$^{14}$. In this approach the FQHE wavefunction is
constructed by attaching an even number of magnetic fluxes to electrons
filling an integer number of Landau levels. The incompressibility of the IQHE
is thus carried over to the FQHE wavefunctions.
As far as the hierarchical filling fractions $\n= {n \over
{2pn+1}}$ are concerned, the essential idea of this scenario is that the
corresponding incompressible FQHE ground state wavefunction $\Psi _{\n} ^0$, is
related to the IQHE wavefunction $\Psi _n ^0$ as
$$ \Psi _{\n} ^0 = \prod _{i<j} (\zbar _i - \zbar _j)^{2p} \Psi _n ^0 \eqno
(2.11)
$$
This relation is a straightforward generalization of Eq. (2.1). Using previous
ideas we find that the corresponding \Winf generators are of the form$^{7}$
$$
W_{2p} ^{I} = \int d^2 z e^{-|z|^2} \psi ^{\dag I} (z, \zbar) e^{2p \sum _J \a
^J (\zbar)} ~ \ddag \xi (\del _{\zbar}, \zbar - \del _z)
\ddag ~ e^{-2p \sum _J \a ^J (\zbar)} \psi ^I (z, \zbar) \eqno (2.12)
$$
where $I=0,...,n-1$, give rise to $n$ commuting copies of \Winf algebras and
the corresponding Jain ground state for $\n ={n
\over {2pn+1}}$ satisfies the highest weight condition.
\vglue 0.6cm
\line{\elevenbf 3. References \hfil}
\vglue 0.4cm
\item{1.} R.E. Prange and S.M. Girvin, {\elevenit The Quantum Hall Effect}
(Springer, NY, 1990).
\item{2.} M. Stone, {\elevenit Quantum Hall Effect} (World
Scientific, 1992).
\item{3.} S. Iso, D. Karabali and B. Sakita, {\elevenit Nucl. Phys.} {\elevenbf
B388} (1992) 700.
\item{4.} S. Iso, D. Karabali and B. Sakita, {\elevenit Phys. Lett.} {\elevenbf
B296} (1992) 143.
\item{5.} A. Capelli, C. Trugenberger and G. Zemba, {\elevenit Nucl. Phys.}
{\elevenbf B396} (1993) 465.
\item{6.} D. Karabali, {\elevenit Nucl. Phys.} {\elevenbf B419} (1994) 437.
\item{7.} D. Karabali, {\elevenit Nucl. Phys.} {\elevenbf B428} (1994) 531.
\item{8.} S.M. Girvin and T. Jach, {\elevenit Phys. Rev.} {\elevenbf B29}
(1983)
5617.
\item{9.} I. Bakas, {\elevenit Phys. Lett.} {\elevenbf
B228} (1989) 57; I. Bakas and E. Kiritsis {\elevenit Nucl. Phys.} {\elevenbf
B343} (1990) 185; D.B. Fairlie and C.K. Zachos, {\elevenit Phys. Lett.}
{\elevenbf B224} (1989) 101; J. Hoppe and P. Schaller,
{\elevenit Phys. Lett.} {\elevenbf B237} (1990) 407.
\item{10.} M. Stone, {\elevenit Phys. Rev.} {\elevenbf B42} (1990) 8399;
{\elevenit Ann. Phys.}
{\elevenbf 207} (1991) 38; for a review on edge excitations see, X.G. Wen,
{\elevenit Int. J. Mod.
Phys.} {\elevenbf B6} (1992) 1711.
\item{11.} A. Cappelli, C.A. Trugenberger and G.R. Zemba, {\elevenit Phys.
Lett.} {\elevenbf B306} (1993) 100; A. Cappelli, G.V. Dunne, C.A. Trugenberger
and G.R. Zemba, {\elevenit Nucl. Phys.} {\elevenbf B398} (1993) 531.
\item{12.} R.B. Laughlin in ref.[1]; {\elevenit Phys. Rev. Lett.} {\elevenbf
50} (1983) 1395.
\item{13.} F.D.M. Haldane in ref. [1]; F.D.M. Haldane, {\elevenit Phys. Rev.
Lett.} {\elevenbf 51} (1983) 605; B.I.
Halperin, {\elevenit Phys. Rev. Lett.} {\elevenbf 52} (1984) 1583.
\item{14.} J.K. Jain, {\elevenit Phys. Rev. Lett.} {\elevenbf 63} (1989) 199;
{\elevenit Phys. Rev.} {\elevenbf B40} (1989) 8079; {\elevenit Phys. Rev.}
{\elevenbf B41} (1990) 8449.
\item{15.} M. Flohr and R. Varnhagen, {\elevenit J. Phys.} {\elevenbf A27}
(1994) 3999.
\item{16.} A. Cappelli, C. Trugenberger and G. Zemba, {\elevenit Phys. Rev.
Lett.}
{\elevenbf 72} (1994) 1902.
\item{17.} S.M. Girvin, {\elevenit Phys. Rev.} {\elevenbf B29} (1984) 155.
\item{18.} N. Read, {\elevenit Phys. Rev. Lett.} {\elevenbf 65} (1990) 1502;
J. Frolich
and A. Zee, {\elevenit Nucl. Phys.} {\elevenbf B354} (1991) 369; X.G.
Wen and A.Zee, {\elevenit Phys. Rev.} {\elevenbf B46} (1992) 2290.

\end